\documentclass[twocolumn,useAMS,usenatbib]{mnras}

\usepackage{amsmath,amssymb}
\usepackage{color}
\usepackage{natbib}
\usepackage{graphicx}

\newcommand{\be}{\begin{equation}}
\newcommand{\ee}{\end{equation}}
\newcommand{\WL}{\mathrm{WL}}
\newcommand{\X}{\mathrm{X}}
\title[Weak Lensing of the RBC]{Weak lensing calibration of mass bias in the REFLEX+BCS X-ray galaxy cluster catalogue}

\author[Simet et al.]{Melanie Simet$^1$, Nicholas Battaglia$^{2,1}$, Rachel Mandelbaum$^1$, Uro\v{s} Seljak$^{3,4}$\\
$^1$McWilliams Center for Cosmology, Carnegie Mellon University, 5000 Forbes Ave, Pittsburgh PA, USA, 15213\\
$^2$Department of Astrophysical Sciences, Princeton University, Princeton, NJ 08544\\
$^3$Physics, Astronomy Department, University of California, Berkeley, California, USA\\
$^4$Lawrence Berkeley National Laboratory, Berkeley, CA, USA}

\begin{document}

\maketitle

\begin{abstract}
The use of large, X-ray selected galaxy cluster catalogues for cosmological analyses requires a thorough understanding of the X-ray mass estimates.
Weak gravitational lensing is an ideal method to shed light on such issues, due to its insensitivity to the cluster dynamical state. We perform a weak lensing calibration of 166 galaxy clusters from the REFLEX and BCS cluster catalogue and compare our results to the X-ray masses based on scaled luminosities from that catalogue.  To interpret the weak lensing signal in terms of cluster masses, we compare the lensing signal to simple theoretical Navarro-Frenk-White models and to simulated cluster lensing profiles, including complications such as cluster substructure,
projected large-scale structure, and Eddington bias.  We find evidence of underestimation in the X-ray masses, as expected, with $\langle M_{\X}/M_{\WL}\rangle = 0.75 \pm 0.07$ stat. $\pm 0.05$ sys. for our best-fit model. The biases in cosmological parameters in a typical cluster abundance measurement that ignores this mass bias will typically exceed the statistical errors.

\end{abstract}


\section{Introduction}

Galaxy clusters are the most massive objects in the universe, reaching masses of $10^{15} M_{\odot}$ or more.  Their abundances are a sensitive probe of the growth of structure because they measure the exponential tail of the halo mass function \citep[e.g.][]{Bahcall88,Vikh2009}.  They also provide an important cosmological probe, as they are complementary to the traditional geometric probes \citep{DarkEnrgTF} and can be used to test modified theories of gravity \citep{2008PhRvD..78f3503J}.  However, for robust interpretation of cluster abundance measurements we must know the masses of the clusters in the sample; uncertainties in inferring cluster masses from observations are a major limitation to the usefulness of cluster abundances in precision cosmological measurements \citep{2009ApJ...692.1060V,2010MNRAS.405.2078M,2011ARA&A..49..409A,2013SSRv..177...75H,2013PhR...530...87W,2014MNRAS.439....2V}.

Cluster mass measurements or mass proxies can be defined from measurements in multiple wavebands, deriving from different components of the cluster structure.  From simulation and observation, we know that the dominant component of galaxy clusters is dark matter, usually distributed in a somewhat clumpy, roughly triaxial structure with a density profile well-described by a Navarro-Frenk-White profile \citep[NFW;][]{1997ApJ...490..493N} or the related Einasto profile \citep{2010MNRAS.402...21N}. The dominant baryonic component of clusters is a reservoir of hot, ionized gas that collects in the potential well of the cluster.  In the X-ray, this gas is clearly visible as a bright, extended extragalactic source.  In the radio, the gas also appears as deviations in the cosmic microwave background, known as the Sunyaev-Zel'dovich effect \citep{SZ1970}: the hot gas Compton upscatters CMB photons, leaving a decrement or increment in the CMB brightness depending on wavelength.  Finally, in the optical, galaxy clusters are characterized by an overdensity of galaxies in a small region of physical and redshift space.  Mass estimates can be made using any of these wavebands as proxies: scalings to optical richness (number of galaxies in the overdensity), strength of the Sunyaev-Zel'dovich effect, and X-ray quantities such as temperature and luminosity have all been used, in addition to the weak lensing measurements discussed below.

Numerical studies indicate that X-ray masses which assume hydrostatic equilibrium are biased low at the 10-30 per cent level depending on assumptions about feedback and preheating \citep[e.g.,][and references therein]{Rasia2006,Lau2009,Batt2012a,Rasia2012,2012MNRAS.422.1999K,2014ApJ...782..107N}. This is generally called the ``hydrostatic mass bias.'' In addition, errors may be introduced through differing X-ray normalizations and measurements.  The total offset between X-ray mass measurements or mass scalings and weak lensing masses is usually measured as $\langle M_{\X}/M_{\WL}\rangle-1$ or as the slope of an $M_{\WL}$ vs $M_{\X}$ relation minus one; we will call these two cases the ``average bias'' and the ``fit bias.'' 

Several groups have found these mass biases to be different for so-called ``relaxed'' and ``unrelaxed'', or ``undisturbed'' and ``disturbed'', clusters. The division is based on the amount of substructure present in the X-ray gas maps: clusters with little substructure are relaxed or undisturbed, those with a high degree of substructure are unrelaxed or disturbed.  Unrelaxed clusters are thought to indicate recent mergers.  The cluster masses in X-ray are determined as hydrostatic masses (a combination of the X-ray temperature and gas or electron densities out to a given aperture, assuming the gas is fully in hydrostatic equilibrium and often that it is spherically symmetric; see e.g. \citealt{2014A&A...564A.129I} for details) or by a calibrated relationship between total mass and a quantity such as luminosity or temperature, which we will call ``mass scalings'' in this work to distinguish them from mass measurements made using multiple observational quantities.  The aperture is defined by the overdensity $\Delta$, the ratio by which the average density inside the aperture is greater than the critical density $\rho_c$ or (more rarely) the mean matter density $\rho_m$.  For clarity these are often called $\Delta_c$ and $\Delta_m$, respectively. Common choices are $\Delta=500$ for X-ray analyses (which typically cannot detect gas very far out from the center of the cluster) and $\Delta=200$ for weak lensing analyses.

The most direct probe of the cluster mass is weak gravitational lensing \citep[for a review, see][]{2001PhR...340..291B,2003ARA&A..41..645R,schneider06,2008ARNPS..58...99H,2010RPPh...73h6901M}, which is independent of the dynamical state of the cluster and depends only on gravity.  By statistically detecting correlations in the images of more distant galaxies whose light has passed the objects of interest, we learn about the gravitational fields of these objects, meaning weak lensing measurements are equally sensitive to dark matter and baryonic matter \citep{2006glsw.conf....1S}.  Galaxy clusters provided the first demonstration of weak lensing \citep{1990ApJ...349L...1T} and measuring galaxy cluster masses, both on an individual basis and in stacks of similar objects, is an active field of research.  However, weak lensing measurements have their own systematics, including projection effects and the effects of nonspherical haloes and cluster substructure \citep[e.g.,][]{2007MNRAS.380..149C,2010ApJ...709..286M,Mene2010,Beck2011,2012A&A...545A..71V}.

Several previous studies have used weak lensing to calibrate easier to measure, but harder to interpret, cluster observables from multiple wavebands.  We focus here on calibrations of X-ray mass scalings, since these measurements have been used as inputs in recent broad cosmological measurements. See, e.g., the \citet{arnd2010} comparison used as input to the measurements of Planck \citep{2014A&A...571A..20P,2014A&A...571A..16P}; the \citealt{arnd2010} relation was also used to generate the X-ray scaled masses in the RBC catalogue.  Unless otherwise noted, all quantities are reported at $\Delta_c=500$.  We report bias here and throughout the paper as 
\begin{equation}
M_{\X}/M_{\WL} = 1-b  \equiv q.
\end{equation}
\begin{itemize}

\item \citet{2014MNRAS.442.1507G} measure weak lensing masses for twelve SZ-selected galaxy clusters from the South Pole Telescope and compare them to results from the literature from Chandra, ROSAT, and XMM-Newton.  The clusters range from redshifts of 0.1 to 0.7, with most between 0.3 and 0.4, and from masses of $0.4-2\times 10^{15} h_{70}^{-1} M_\odot$.  They find results for a fit bias consistent with no hydrostatic mass bias ($2_{-15}^{+10}$ per cent) but also marginally consistent with the expected $\approx 20\%$ bias from simulation. 
 
\item In an analysis of four relaxed clusters with joint Suzaku and Subaru observations, \citet{2014arXiv1406.3451O} compute hydrostatic masses and find a $\sim 24\pm 7\%$ average bias at $\Delta_c=500$, increasing to $\sim 60\pm 10\%$ at the virial overdensity.  Redshifts range from $0.05$ to $0.25$ and masses (in this case $M_{200,c}$) from $0.4-1.7\times 10^{15} h_{70}^{-1} M_\odot$.  They, too, observe increased bias at larger radii.

\item \citet{2014A&A...564A.129I} examine 8 clusters from the {\it 400d} survey.  The clusters have redshifts $0.39<z<0.8$ and masses $1-6\times10^{14} M_\odot$.  They compare weak lensing masses with hydrostatic mass estimates and find results consistent with no bias, and with somewhat lower scatter than expected.  

\item \citet{2014arXiv1405.7876D} investigate 25 CLASH clusters using weak and strong lensing profiles in conjunction with XMM-Newton and Chandra hydrostatic profiles combined as in \citet{2013ApJ...767..116M}. The CLASH clusters range from $0.2<z<0.9$ and masses $M_{2500,c}$ from $1-10\times10^{14} h_{70}^{-1} M_\odot$.  \citeauthor{2014arXiv1405.7876D} find an average bias, but the magnitude and sign of the bias depends on the analysis type and the aperture used for the measurement.  For the 0.5 Mpc measurement, the most similar to our measurements here, they find an average bias of $12 \pm 7$ per cent for weak lensing (consistently with previous measurements) but $-11 \pm 7$ for a strong and weak lensing combined measurement.  The average bias increases to 20 to 30 per cent at 1 Mpc.

\item \citet{2015arXiv150902162A} compare twelve relaxed clusters with X-ray data from Chandra
  with weak lensing masses from the Weighing the Giants programme.  The clusters have redshifts
  $0.15-0.54$ and masses $M_{2500,c}$ from $2-12\times10^{14} M_\odot$.  The average bias is
  found to be $-4 \pm 8$ per cent, with the weak lensing masses smaller than the X-ray masses
  but consistent within the errors.  The same group made a measurement \citep{2014MNRAS.443.1973V} using this set of twelve clusters plus ten others and compared the weak lensing measurements to SZ masses instead of X-ray masses, finding a bias of $31 \pm 7$ per cent, with the SZ masses smaller than the weak lensing masses. The authors attribute these differences to differences in the X-ray calibration between Chandra and XMM, which has been seen in other studies as well.
\end{itemize}

Some earlier studies were performed with less constraining power, such as
\citet{2012A&A...546A.106F} from EXCPRES and \citet{2007A&A...467..437Z, 2008A&A...482..451Z,
  2010ApJ...711.1033Z} from LoCuSS.  Comparisons are also performed against other X-ray
  measurements, such as temperature; for example, \cite{2015MNRAS.448..814I} find a bias of about 35
  per cent comparing to temperature, but note that this is different by 15-20 per cent than the bias
  measured with \textit{Chandra} instead of \textit{XMM-Newton}, pointing to the influence of X-ray
  calibrations on the bias. Other studies \citep[e.g.][]{2014MNRAS.443.1973V} have calibrated masses from cosmic microwave background data, which are ultimately scaled from X-ray masses, but are not a direct comparison.  Still others \citep[e.g.][]{2014arXiv1402.4484G, 2013ApJ...767..116M} fit the X-ray scalings directly with weak lensing data, without first passing through a theoretically motivated proxy-mass relationship.

In this paper, we use weak lensing to measure the masses of a subsample of the MCXC X-ray cluster
catalogue \citep{piff2011} called the REFLEX+BCS+CIZA~(RBC) catalogue \cite[e.g.,][]{HB2013}, which
has a well-defined selection function with a set of predicted masses from scaled X-ray luminosities
\citep{arnd2010}. (We note that due to extinction cuts $A_r<0.2$ for the shear catalogue, we do not include any CIZA clusters, but we retain the acronym since the catalogue has a common format and standardized measurements.)  We recalibrate these masses using the subset of RBC clusters that are found in the region of the Sloan Digital Sky Survey \citep[SDSS;][]{2000AJ....120.1579Y}, using the weak lensing signal measured using background galaxy shapes and photometric redshifts from SDSS.  We also have a set of hydrodynamic simulations with similar characteristics to the RBC catalogue, which we will exploit to produce a model for the weak lensing signal including non-trivially complex physical effects such as triaxiality and substructure.  As the RBC is an all-sky survey, and thus overlaps with many upcoming surveys and smaller projects, we hope this calibration will be useful for many groups.

We cover the theoretical background of weak lensing in Section \ref{Methodology}.  Section \ref{Data} describes our data and simulations, and Section \ref{Results} shows the results of our measurements.  We summarize the mass calibration in Section \ref{Conclusion}.

\section{Methodology}\label{Methodology}

\subsection{Weak lensing}

Weak gravitational lensing leads to a correlation between the shapes of galaxies due to the gravitational influence of the matter between those objects and the observer.  A concentrated overdensity of matter, such as a cluster, will alter the apparent shape of the background galaxies to produce tangential correlations: due to the gravitational field of the overdensity, the average galaxy shape in an annulus will appear to lie preferentially along the annulus itself, rather than randomly or along the radial direction.  We call this distortion of the shape a shear.  Weak gravitational lensing is called ``weak'' because these shears are small.  

We work in the thin-lens limit, where the line-of-sight extent of the lens is much smaller than both the distance between the observer and the lens and the distance between the lens and the background galaxy.  A lens with surface mass density $\Sigma(R)$ at redshift $z_l$ will induce on a background galaxy at redshift $z$ a tangential shear $\gamma_t$~\citep{2006glsw.conf....1S}:
\begin{equation}
\gamma_t(R) = \frac{\bar{\Sigma}(<R) - \bar{\Sigma}(R)}{\Sigma_{\mathrm{cr}}(z_l,z)}.
\end{equation}
The shear is proportional to the difference between the average surface mass density interior to radius $R$ ($\bar\Sigma(<R)$) and the average surface mass density at the radius $R$ ($\bar{\Sigma}(R)$).  No assumption of spherical symmetry is necessary: the equation is exact for any overdensity (or underdensity) of matter as long as the average around the entire annulus or disk is considered.  
The shear is also inversely proportional to the critical surface mass density, or $\Sigma_{\mathrm{cr}}$, which is given by a scaled combination of angular diameter distances:
\begin{equation}
\Sigma_{\mathrm{cr}} = \frac{c^2}{4\pi G} \frac{D_A(z)}{(1+z_l)^{-2}D_A(z_l)D_A(z_l,z)}
\end{equation}
where the extra factor of $(1+z_l)^{-2}$ comes from our use of comoving coordinates.
A more formal definition of weak lensing, as opposed to strong lensing, is the regime where $\Sigma(R)/\Sigma_{\mathrm{cr}}(R) \ll 1$.

We combine the noisy per-object galaxy shape measurements into a single statistic via an estimator
\begin{equation}\label{DeltaSigmaEstimator}
\widetilde{\Delta\Sigma} = C(R) \frac{\sum_i w_i \Sigma_{\mathrm{cr},i} \gamma_{t,i}}{\sum_i w_i}.
\end{equation}
The optimal weighting for this equation includes both the per-object shape noise weighting $w_{i,\mathrm{shape}}$ and the critical surface mass density:
\begin{equation}\label{Weights}
w_i = \Sigma_{\mathrm{cr}}^{-2} w_{i,\mathrm{shape}}
\end{equation}
\citep{2007arXiv0709.1159J, 2012ApJ...757....2G}.  The critical surface mass density optimally reduces the scatter caused by the different lensing geometries of background galaxies at different redshifts.  The per-object shape noise weighting will be described in Section \ref{WLdata}.

The optimal weighting also includes what is known as a boost factor correction, $C(R)$. This
accounts for the contamination of the background galaxy sample by galaxies associated with the
clusters themselves, which have a shear expectation value of 0, rather than the positive value
expected for lensed galaxies.  Usually this correction is made via a comparison to weighted number
densities (using the weights in Eq.~\ref{Weights}) around random points distributed in the same way
as the lenses \citep{2004AJ....127.2544S}.  

For our purposes, with such a small number of lenses
covering the full extent of the somewhat heterogeneous background galaxy sample, random points would
likely not produce the expected distribution of background galaxy densities.  Instead, in this
  analysis we take the average density in a large-scale annulus around the clusters -- from
$18$ to $40 h_{70}^{-1}$ Mpc -- which is larger than our field of interest and unlikely to be
contaminated by cluster galaxies, but is still physically close enough on the sky to probe the same
field number densities as the lenses themselves.  Since survey edges intersect this largest
  annulus (and, to some degree, the annuli closer to the cluster centers), we cannot use the
  expected area of the annulus to modify the galaxy densities.  Rather, we measure the number of red galaxies with photometric redshift above 0.5.  For red galaxies at higher redshift, our photometric redshift estimates are secure and contamination from cluster member galaxies is negligible.  This sample is too small to detect the lensing effect above the
  noise, but it is large enough to be suitable for measuring the source number density.  For a constant density of red galaxies, a fair assumption for the SDSS catalogue, the area is simply proportional to the number count of red galaxies in the annulus.  This gives us a boost factor
\begin{equation}\label{BoostFactorEquation}
C(R_1 < R < R_2) = \left(\frac{N_{\mathrm{red}}(18 < R_j < 40)}{N_{\mathrm{red}}(R_1<R<R_2)}\right)\frac{\sum_{i: R_1<R<R_2} w_i}{\sum_{j: 18 < R_j < 40} w_j}
\end{equation}.

\begin{figure}
\includegraphics[width=0.45\textwidth, clip, trim=0.7in 2in 0.3in 1in]{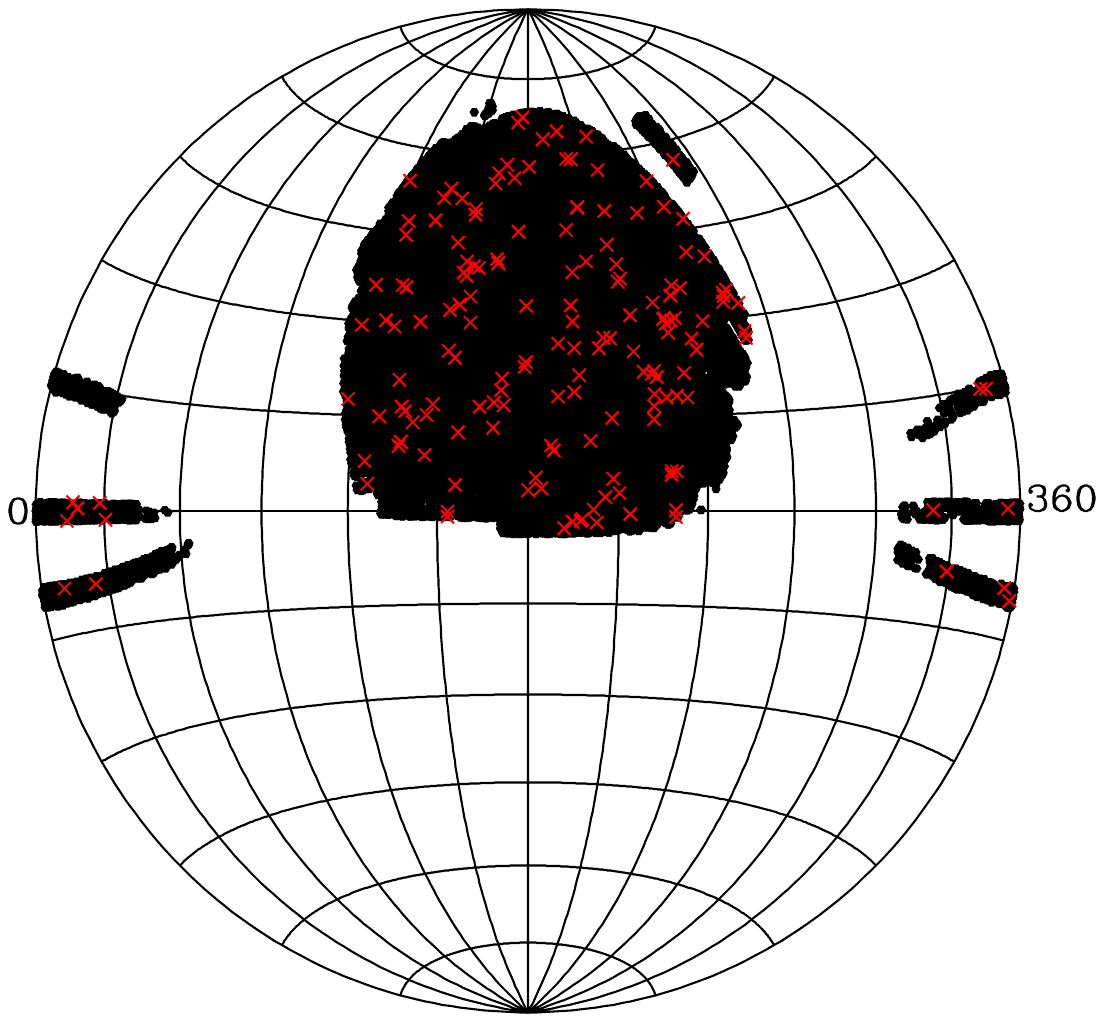}
\caption{Sky coverage of the SDSS shape catalogue (in black) and the overlapping RBC clusters (red Xs).}\label{coverage}
\end{figure}

Because of the extreme contamination due to survey edges in the sparsely-covered southern Galactic cap region of SDSS (the three stripes broken at a right ascension of 0 in Fig. \ref{coverage}), even this method was insufficient to accurately measure the boost factor in that region.  The boost factor should be consistent for clusters of the same richness and redshift, so we should be able to use the 150 clusters of the northern cap to stand in for the 16 southern cap clusters if the richness and redshift distributions are consistent.  We find that the richness distribution is consistent, but the redshift distribution shows a larger concentration of low-redshift clusters compared to the northern cap.  We at first attempted measuring the boost factor in redshift slices and reweighting to obtain the global boost factor, but we found that the reweighted boost factors had approximately the same errors relative to the all-redshift measurement as a simple unreweighted combination of the redshift slices, so we use the directly measured boost factor for clusters of all redshifts in the northern cap as the boost factor for the entire sample.  This result holds when we bin the clusters in X-ray mass.

Errors on the signal are estimated as the standard deviation of the mean for each sum in Equations \ref{DeltaSigmaEstimator} and \ref{BoostFactorEquation}.  Since we are looking here at small scales around widely-separated sources, we do not expect a noticeable covariance between adjacent radial bins.

The $\Delta\Sigma$ measurements as a function of radius give us an estimate of the 2-dimensional surface mass distribution in the region of the cluster, which we must turn into a 3-dimensional mass contained within some region of interest.  There are nonparametric methods of performing this inversion, but since we believe we know the density profiles of galaxy clusters, interpretation is made simpler by using preexisting 3-dimensional models projected down to 2-dimensional surface mass densities, which we can then compare to our measured $\Delta\Sigma$ profiles.  In this work, we will compare our results with theoretical NFW profiles and with results from hydrodynamic simulations described in Section \ref{sec:sim}.

\subsection{Miscentering}\label{sec:mis_cent}

We measure the weak lensing signal relative to the reported galaxy cluster center, which is the X-ray centroid from the X-ray catalogue described in Section \ref{Data-RBC}.  Errors in the determination of the cluster center will suppress the weak lensing signal on small scales.  Since we are taking averages of shears in annuli, an annulus around a location other than the peak will contain a small region of higher shear than expected and a much larger region of lower shear than expected.  This effect reduces the average shear in the annulus, and the effect size is greater at smaller radii, where the shear is changing rapidly with radius.  If not accounted for directly, this suppression will cause errors in our final mass measurement \citep{2007arXiv0709.1159J,2008JCAP...08..006M,2011ApJ...740...53R} that can be tens of per cent, comparable to the size of the mass bias that we are trying to measure.   

Errors in centroid determination can come from physical sources (unrelaxed clusters whose center of mass may be physically separated from the X-ray peak) and from measurement uncertainty (the finite resolution of the X-ray map).  For optical cluster catalogues, some fraction of the clusters are exactly centered, due to the exact centering of some brightest cluster galaxies. The rest have a distribution of $R_s$, the distance between the true center and the observed center, usually written as a Rayleigh distribution \citep{2007arXiv0709.1159J}
\be
P(R_s) = \frac{R_s}{\sigma_s^2} \exp\left(-\frac{1}{2}(R_s/\sigma_s)^2\right)
\label{eq:mis_cent}
\ee
which is the average absolute offset for a measurement with Gaussian error in both the $x$ and $y$ directions.  For the X-ray measurements, we have no perfectly centered clusters since centroid measurement uncertainty affects them all, so we may use $P(R_s)$ directly for our sample as a whole.  For $\sigma_s$, the width of the distribution of centroiding errors, we use both $0.25\arcmin$ and $0.5\arcmin$, related to the average resolution of the X-ray maps used in the creation of the RBC catalogue described in section \ref{Data-RBC}; this is also consistent with the distribution of centroid differences between CHANDRA and ROSAT Abell clusters, as discussed in \citet{2014ApJ...797...82L}.  We also tested several other values, as discussed in section \ref{Results}, but these values encompass the best fits to the data, as described further in section~\ref{Results}.  The cluster offsets are a combination of true offsets between X-ray centers and dark matter centers, plus measurement error in the X-ray center; for clusters in this redshift range and with the given X-ray precision, the measurement error is the larger consideration, so we use it as our main proxy.  Our theoretical mass profiles $\Sigma(R)$ are effectively convolved with this distribution of centroid offsets before calculating the theoretical lensing signals, $\Delta\Sigma$.  For our simulation results, this convolution is done on a cluster-by-cluster basis using the simulation redshifts, since we must convert from angular to physical coordinates.  For the NFW profiles, due to computation time limits, the clusters are grouped into redshift slices containing 10-20 clusters, sometimes less at the high-redshift end of the sample. The average weight and angular diameter distance of the groups are calculated using weights based on the lensing geometry, and a weighted summed profile is constructed using the results from each group of clusters instead of each cluster individually.

\section{Data and Simulations}\label{Data}

\subsection{The RBC cluster sample}\label{Data-RBC}

The cluster catalogue that we used is the RBC catalogue, which is a sub-sample of the MCXC X-ray cluster catalogue \citep{piff2011}. This sub-sample combines the REFLEX \citep{bohr2004}, BCS \citep{ebel1998,ebel2000} and CIZA \citep{ebel2002,koce2007} flux limited catalogues from the RoSAT All Sky Survey \citep[RASS;][]{voge1999}. (We do not include any CIZA clusters after our lensing cuts--see below for more information.)  We assume the RBC catalogue is complete above a flux limit of $\sim 3\times 10^{-12}/ \rmn{erg}\,\rmn{s}^{-1} \rmn{cm}^{-2}$, or $4.4 \times 10^{-12}/ \rmn{erg}\,\rmn{s}^{-1} \rmn{cm}^{-2}$ for the BC subsample, so we look at clusters only above this flux limit.  These are the same cuts used in \citet{Mantz2010}, which also contains more information on possible incompletenesses. We consider only the 166 clusters that overlap with the SDSS lensing footprint. The cluster masses are taken from the MCXC catalogue, which uses a $L_\X-M$ scaling relation (Malmquist bias corrected) from the REXCESS cluster sample \citep{prat2009}. The scaling relation is

\be \label{eq:lxm}
h(z)^{-7/3} \left(\frac{L_{500}}{10^{44}\rmn{erg}\,\rmn{s}^{-1}}\right) = C\left(\frac{M_{500}}{2 \times 10^{14} M_{\odot}}\right)^{1.64}
\ee
where $h(z) \equiv [\Omega_\rmn{M}(1+z)^3 + \Omega_{\Lambda}]^{1/2}$ and log$(C) = 0.274$. Additionally, there is scatter about this relation of $\sim 24$\% in mass \citep[or $\sim 38$ \% in $L_\X$][]{prat2009}. The scaled cluster masses in the \citet{prat2009} relation were initially calculated under the assumption of hydrostatic equilibrium \citep[HSE][]{point2005} for a sample of 10 relaxed clusters. They were then used to calibrate the $Y_\X-M$ \citep{krav2006} relation in \citet{arnd2007} and updated in \citet{arnd2010}. This larger set of clusters with a $Y_{\X}-M$ relation were then used to infer a relationship between $M$ and $L_{\X}$.  This scaling relation was then extended to an even larger sample of clusters in the analysis of \citet{prat2009}. Therefore, the assumption of HSE is still present in the masses presented in the MCXC catalogue, albeit at some remove. We do not account for an HSE bias or any additional bias that would result from using this $L_\X-M$ relation, since that bias is what we are trying to measure.

\begin{figure}
\includegraphics[width=0.45\textwidth, clip]{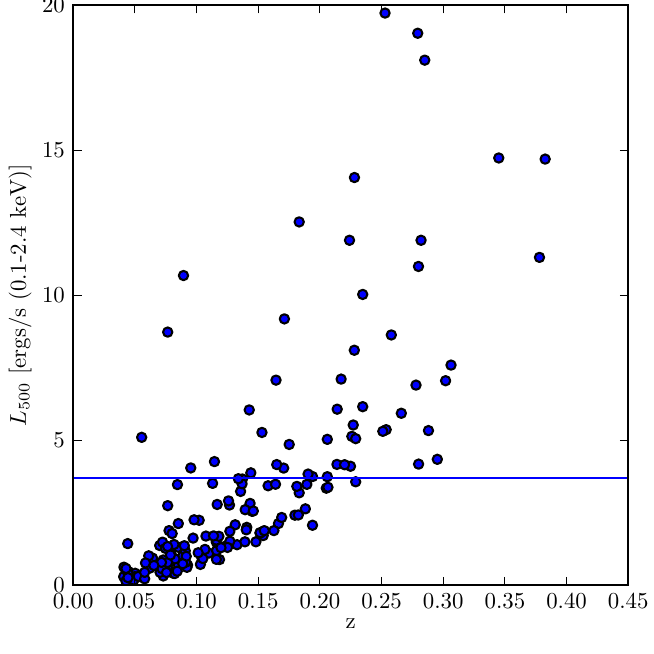}
\caption{Luminosity and redshift coverage of the RBC clusters that overlap with the SDSS source
    catalogue used in this paper.  The horizontal line indicates the approximate location of the split between our two mass bins.}\label{lum-z}
\end{figure}

The cluster sample in the SDSS region ranges from $z\sim 0.04 - 0.4$ with a mean redshift of $0.14$.
The reported masses from scaled luminosities range from $M_{500} \sim 0.6 - 10.5 \times
10^{14} h_{70}^{-1} M_\odot$, with a mean of $3.4 \times 10^{14} h_{70}^{-1} M_\odot$.  A map of the
SDSS coverage area and the overlapping RBC clusters is shown in Fig. \ref{coverage}, and the
catalogue's redshift and luminosity distribution are shown in Fig. \ref{lum-z}.  We note that no
CIZA clusters appear in the masked catalogue, since they are located in regions of high extinction
$A_r>0.2$
where photometric redshifts cannot be reliably determined.

\subsection{Weak lensing data}\label{WLdata}
\subsubsection{Galaxy shape sample}

We use a shape catalogue measured using the re-Gaussianization software \citep{2003MNRAS.343..459H}
applied to data from the Sloan Digital Sky Survey (SDSS) data release 8
\citep{2000AJ....120.1579Y,2011ApJS..193...29A}.  This catalogue is described in detail in
\citet{2012MNRAS.425.2610R}, including details of shear calibration and error estimation.  The
software calculates adaptive second-order moments for the galaxy images and PSFs by effectively
fitting them to elliptical Gaussians, and those moments are then combined to generate PSF-corrected
distortion measurements including a correction for low-order non-Gaussianity of the galaxies and
PSFs.  We turn the distortions into shears by dividing by a shear responsivity factor $\cal R$
\citep{2002AJ....123..583B} which for our shear catalogue is approximately 1.8
\citep{2012MNRAS.425.2610R}. This factor describes the average sensitivity of the shape sample to an
applied shear, as shears do not add linearly \citep{2002AJ....123..583B}.  After cuts removing badly
measured or unmeasurable objects, the catalogue contains 39 million galaxies in an area of
approximately 9,000 square degrees, for an overall unweighted number density of 1.2
galaxies/arcmin$^2$.  This shear catalogue was tested extensively in
  \citet{2012MNRAS.425.2610R}.  Additionally, further testing was done using simulated images, which
  found that combined errors from PSF correction, noise bias, and selection bias gave a systematic
  shape error budget of 3.5 per cent \citep{2012MNRAS.420.1518M}; combined with other calibration
  errors such as photo-$z$ error and stellar contamination, $\Delta\Sigma$ profiles calculated using
  this shape sample should have a total systematic error budget of 4 per cent including all shear
  and photo-$z$ biases \citep{2013MNRAS.432.1544M}, below the statistical error of our measurement.

The galaxies have a per-object shape weight of
\begin{equation}\label{WeightEquation}
w_{i,\mathrm{shape}} = \frac{1}{e_{\mathrm{rms}}^2 + \sigma_{e,i}^2}.
\end{equation}
which is a combination of the individual measurement error $\sigma_{e,i}$ and the intrinsic dispersion of galaxy shapes $e_{\mathrm{rms}}$, which is 0.36 for this sample.

\subsubsection{Galaxy redshifts}\label{WLphotoz}

The galaxy photometric redshifts are presented in \citet{2012MNRAS.420.3240N}.  The measurements were made with the Zurich Extragalactic Bayesian Redshift Analyzer, or ZEBRA, a template-fitting software~\citep{2006MNRAS.372..565F}.  Thirty-one templates were used, with six templates (four observed SEDs and two synthetic blue galaxy spectra) from \citet{2000ApJ...536..571B} and twenty-five additional templates created via interpolation between neighboring pairs of the six starting templates.  For this catalogue, ZEBRA was run in a mode that produced maximum-likelihood point estimates for the photometric redshifts, and not $p(z)$ distributions.   Starburst-type galaxies, representing 10 per cent of the sample, were removed due to unreliability after comparison against spectroscopic and high-quality photometric redshifts.  The remaining galaxies have a known bias for $z \gtrsim 0.4$, approximately the mean redshift of the catalogue.  The bias in the lensing signal due to background galaxy photo-$z$ bias and scatter can be corrected since we know the true $dN/dz$  \citep{2012MNRAS.420.3240N}.  For this cluster sample, in which the majority of galaxy cluster lenses are at $z\lesssim 0.1$, the typical correction is small, from 0 to 5 per cent depending on binning scheme.

The noise in the photo-$z$ estimator also makes our optimal weighting, Eq.~\eqref{Weights}, somewhat less optimal, corresponding to an approximately 10 per cent increase in the error bars compared to a case with no noise in the source galaxy redshifts \citep{2012MNRAS.420.3240N,2008MNRAS.386..781M}.  However, this is still a factor of $\sim 2$ better than the case without optimal weighting.

\subsection{Cosmological Hydrodynamic Simulations}
\label{sec:sim}

We used simulations of galaxy clusters to interpret the weak lensing observations. The clusters were extracted from simulated cosmological volumes ($L=165$ Mpc$/h$) such that there was a  sufficient statistical sample of clusters to calculate the lensing signal.  These simulations were run with a modified version of the GADGET-2  smoothed particle hydrodynamics (SPH) code \citep{Gadget}. This version of the GADGET-2 code included sub-grid models for active galactic nuclei (AGN) feedback \citep[for more details see][]{Batt2010}, radiative cooling, star formation, galactic winds, supernova feedback \citep[for more details see][]{SpHr2003}, and cosmic ray physics \citep[for more details see][]{2006MNRAS.367..113P,2007A&A...473...41E,2008A&A...481...33J}. We ran 10 165 Mpc$/h$ boxes with a resolution of 256$^3$ gas and dark matter (DM) particles which yields a mass resolution of $M_\rmn{gas} = 3.2\times 10^{9} \rmn{M}_{\odot}/h$ and $M_\rmn{DM} = 1.54\times 10^{10} \rmn{M}_{\odot} /h$. The cosmological parameters used for these simulations were $\Omega_\rmn{M} = \Omega_\rmn{DM} + \Omega_\rmn{b} = 0.25$, $\Omega_\rmn{b} = 0.043$, $\Omega_\Lambda = 0.75$, $H_0=100\,h\,\rmn{km}\,\rmn{s}^{-1}\,\rmn{Mpc}^{-1}$, $h=0.72$, $n_\rmn{s} =0.96$ and $\sigma_8 = 0.8$.  The outputs of the simulation are in comoving coordinates.

At each redshift snapshot the halos were identified and their properties were calculated in two steps:
\begin{itemize}
\item The halos were found using a friends of friends algorithm \citep{Huch1982}. 
\item For each halo the center of mass was computed iteratively, and then the spherical overdensity mass ($M_{\Delta}$) and radius ($R_{\Delta}$), where $\Delta$ refers to the multiplicative factor applied to the critical density, $\rho_\rmn{cr}(z) \equiv 3 H_0^2 [\Omega_\rmn{M}(1+z)^3 + \Omega_{\Lambda}] / (8\pi\,G)$. 
\end{itemize}
For each redshift slice we project the mass distributions of the halos.
These projections are $8 \times 8$ Mpc$/h$ comoving centered on the cluster's COM. We made separate gas, stellar, and DM components projected down the entire 165 Mpc$/h$ length. Then we summed the projections of all three components for our analyses. 

The simulated stacked $\Delta \Sigma$ profiles that we compare to the RBC weak lensing observations are calculated in the following way. We apply a selection function to the simulated halos which uses the scaled masses and redshifts from the RBC catalogue, since the clusters in the RBC sample are flux limited. We model this selection using the initial spherical overdensity masses from the simulations and then we apply a Monte Carlo method that scales each of these masses by a randomly selected number drawn from a Gaussian distribution with a mean of 1 and a dispersion that matches the $L_\X-M$  scatter in mass of 24\%. This correction models the Eddington bias resulting from the scatter in the scaling relation used to calculate the masses in the RBC catalogue. These modeled {\it observed} masses are used to determine which halos are placed in a given mass bin. We then apply the redshift cut, which is well determined for a given mass from the RBC selection function.  We emphasize that we do not ever directly predict luminosity quantities: we always work in terms of mass for these predictions.

For each halo we compute the $\Delta \Sigma$ profile with miscentering included. The probability distribution for miscentering is described by  Eq. \ref{eq:mis_cent}. In the simulations this miscentering is modeled by a Monte Carlo method.  For each cluster we displaced the center of the cluster by

\be
R_s = \sigma_s \sqrt{-2\,\ln{(1-x_\rmn{rand})}}
\label{eq:cent_disp}
\ee

\noindent where $x_\rmn{rand}$ is a random draw from a uniform distribution. Eq.~\ref{eq:cent_disp} results from integrating Eq.~\ref{eq:mis_cent} to get the cumulative distribution function and solving for $R_s$. We used values of $0.25\arcmin$ and $0.5\arcmin$ for $\sigma_s$, and the motivation for this choice is discussed in Sec.~\ref{Results}. For each halo we found a {\it new} center 10 times and then calculated the average $\Delta \Sigma$ profile given these 10 new centers.

The final simulated $\Delta \Sigma$ profiles for a given mass were calculated by averaging over all halos within a given mass range. This includes halos at different redshift outputs in the simulations. We calculate a weighted final average across the redshift outputs, which accounts for both the weak lensing geometry-related weights and the volume factor (comoving distance squared) associated with including multiple redshifts. 

\subsubsection{Eddington and Malmquist bias}

When measuring the $L_\X-M$ relation in real data, the initial catalogue corrected for the Malmquist bias of their sample as follows.  The calibration was measured on an initial sample, and then the relation was altered using simulation to represent what the calibration would have been if no Malmquist bias was present in the observation itself.  This procedure yields a relation which gives a correct mean mass estimate for each $L_\X$ measurement.  

However, once we start to bin in the masses generated from the $L_\X-M$ relation, we introduce Eddington bias due to the scatter in each individual mass.  We model this effect by adding a random number drawn from a distribution with the expected scatter to the known true masses of our simulated clusters before binning, as described above; this mimics the effect of binning the scattered (but no longer Malmquist-biased) masses of the original catalogue.

\section{Results and Discussion}\label{Results}

We show results for the cluster sample as a whole and also for two mass bins, divided as shown in Table \ref{bintable}.  The cluster sample as a whole has the smallest statistical error, since it contains the most objects, with a total of 1.2 million lens-source pairs; however, since it spans more than an order of magnitude in mass, interpretation of results such as a mean mass can be difficult.  But making our bins too narrow will also result in systematic errors, since we are relying on the advantage of stacking many objects to reduce the effects of stochastic uncorrelated large-scale structure and triaxiality.  Even in the two-bin case, our higher-mass bin has fewer than the 100 clusters suggested by \citet{2009MNRAS.396..315C} to effectively reduce triaxial bias, and a similar effect is likely to come into play for other complications whose impact we expect to be reduced by stacking, such as large-scale structure.  Therefore, we do not believe results from smaller bins will be reliable. As stated previously, the sample as a whole is a flux-selected sample (in the X-ray flux), which means that the mean mass is higher at high redshift than at low redshift. However, when we split into two mass bins, the higher mass bin has a high enough mass cut that it is more similar to a volume-limited sample than a flux-limited sample, as shown in Fig. \ref{lum-z}.  We also note that the lower mass bin has no clusters at the higher end of our redshift range.

\begin{figure*}
\includegraphics[width=0.9\textwidth, clip=true, trim=0in 0in 0.7in 0in]{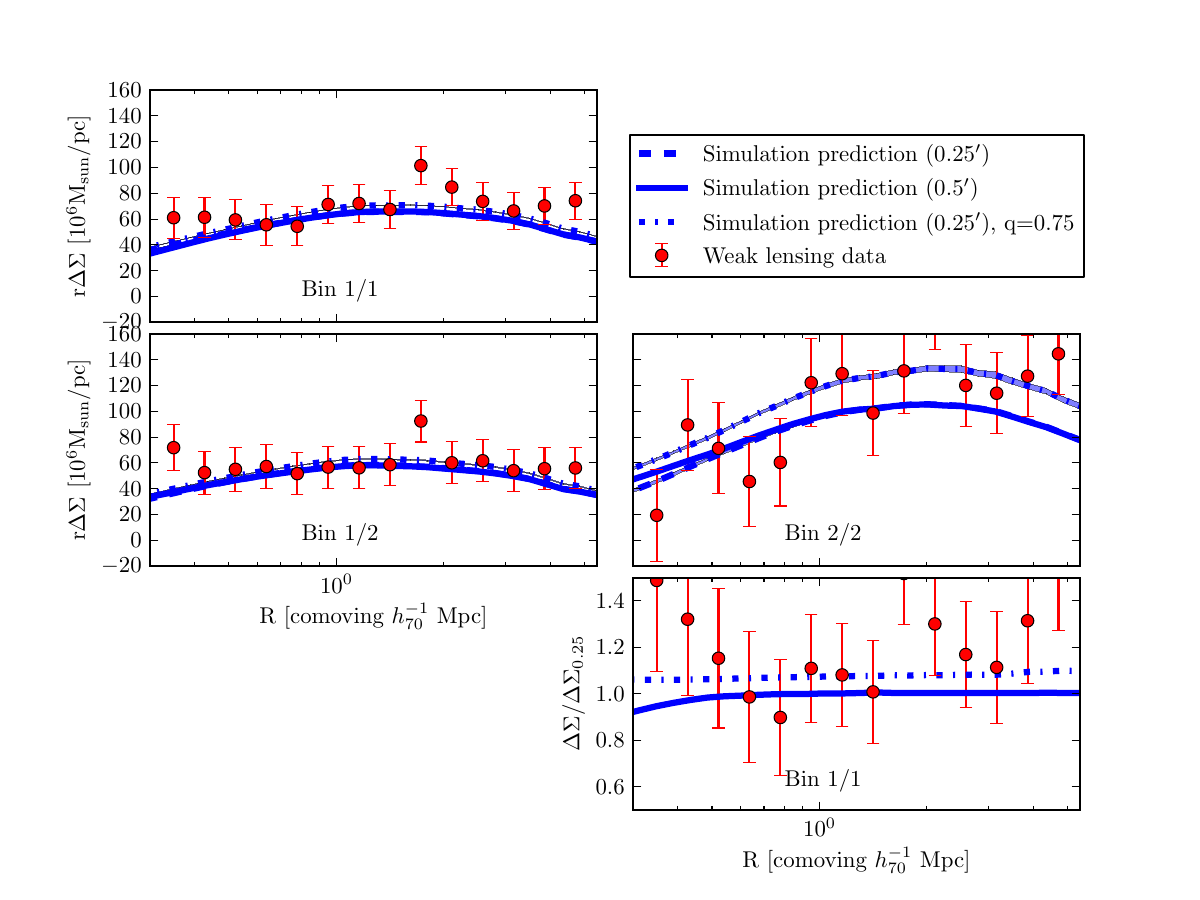}
\caption{Weak lensing measurements (in red) and predicted signal from simulations (solid line for the 0.25\arcmin offset, dashed line for the 0.5\arcmin offset, and dot-dashed for the 0.25\arcmin offset with a calibration of $q=0.75$) for the RBC clusters in two binning schemes with bins split by X-ray mass from scaled luminosities.  Errors are statistical errors on the mean signal for both data sets, using $\sqrt{n}$ scaling.  The bottom right panel shows the ratio between the various curves and the 0.25\arcmin offset curve for Bin 1/1 only, to better highlight the changes caused by these effects.}\label{fig-simpredict}
\end{figure*}

In Fig. \ref{fig-simpredict}, we show the weak lensing signal around the RBC clusters in one or two bins split by X-ray mass as described in Table \ref{bintable}, in addition to the expected signal from the simulations if there is no mass bias ($q=1$) and if there is a 25 per cent mass bias ($q=0.75$).  We expect a slight decrement between the data and the model due to the mass bias (from assumptions of HSE as well as any calibration problems or other sources of error), assuming the cosmology in the simulations is correct; the predictions are low compared to the data, though the difference is not large enough to be conclusive by eye.  We show both the $0.25\arcmin$ miscentering offset and $0.5\arcmin$ offset, since both provide reasonable agreement with the data, as we discuss later in this section.  The ratios between the various predictions or data and the $0.25\arcmin$ offset case are also shown for Bin 1/1 to facilitate comparison.  The $\chi^2$ value for the $q=0.75$ case in Bin 1/1 is 12.0, compared with 18.1 for $q=1$ with the same miscentering offset, showing the clear need for such a calibration assuming the simulations have the correct cosmology. There is no fitting involved in this plot: this is a simple prediction from clusters selected from the simulation using the same selection function as the data.  

\begin{table*}
\centering
\begin{tabular}{| l | c c c c | c | c c c | c c c |}
\hline
    &                   & Mass   & Average       & MB-corrected  & Centroid    & Simulation-fit      &              & \textbf{Mass}                   & NFW  
    &              & \textbf{Mass} \\
Bin & $N_{\textrm{cl}}$ & range  & $M_{\X}$      & $M_{\X}$      & offset      & $M_{\WL}$           & $\chi^2/$dof & \textbf{ratio}                  & $M_{\WL}$     &     $\chi^2/$dof & \textbf{ratio}  \\
\hline
1/1 & 166               & 0.6-15 & $2.8 \pm 0.1$ & $2.5 \pm 0.1$ & 0.5\arcmin  & $3.3 \pm 0.3$       & 8.47/14      & $\mathbf{0.75 \pm 0.07}$        & $3.3 \pm 0.3$ &     8.11/11      & $\mathbf{0.75^{+0.07}_{-0.08}}$ \\
    &                   &        &               &               & 0.25\arcmin & $3.3_\pm 0.3$       & 8.03/14      & $\mathbf{0.75 \pm 0.07}$        & $3.3 \pm 0.3$ & 8.44/11 & $\mathbf{0.77^{+0.07}_{-0.08}}$ \\
\hline
1/2 & 116               & 0.6-4  & $2.1 \pm 0.1$ & $1.9 \pm 0.1$ & 0.5\arcmin  & $2.3 \pm 0.3$       & 7.73/14      & $\mathbf{0.85 \pm 0.01}$        & $2.4^{+0.2}_{-0.3}$ &     6.88/11      & $\mathbf{0.79_{-0.10}^{+0.08}}$\\
    &                   &        &               &               & 0.25\arcmin & $2.3 \pm 0.3 $      & 7.42/14      & $\mathbf{0.84_{-0.09}^{+0.10}}$  & $2.4^{+0.2}_{-0.3}$ & 6.75/11 & $\mathbf{0.81_{-0.10}^{+0.09}}$\\
2/2 & 50                & 4-15   & $5.6 \pm 0.2$ & $4.8 \pm 0.2$ & 0.5\arcmin  & $6.6_{-0.7}^{+0.6}$ & 13.4/14      & $\mathbf{0.73 \pm 0.07}$        & $6.8^{+0.8}_{-1.1}$ &     10.2/11      & $\mathbf{0.71_{-0.07}^{0.12}}$ \\
    &                   &        &               &               & 0.25\arcmin & $6.6_{-0.7}^{+0.6}$ & 12.1/14      & $\mathbf{0.74_{-0.07}^{+0.08}}$ & $6.4^{+0.8}_{-1.1}$ & 11.8/11      & $\mathbf{0.76_{-0.07}^{0.13}}$ \\
\hline
\end{tabular}
\caption{Binning scheme for the RBC clusters, with characteristics of the X-ray masses from scaled luminosities (both the raw average and the average once we have corrected for Malmquist bias), masses fit from the weak lensing data using two lensing models (theoretical NFW and results from hydrodynamic simulations) for two different miscentering parameters, goodness-of-fit statistics, and the resulting X-ray to weak lensing mass ratios.  All masses are in units of $10^{14} h_{70}^{-1} M_\odot$ and are lensing-weighted mean masses, rather than true mean masses; our results for true mean masses are consistent, but require larger corrections to obtain, so we show only the more reliable lensing-weighted averages.  All masses in Bin 1/1 and Bin 1/2 have an additional 5 per cent systematic error bar, and all masses in Bin 2/2 have an additional 17 per cent systematic error bar.}\label{bintable}
\end{table*}

\begin{figure*}
\includegraphics[width=0.9\textwidth, clip=true, trim=0in 0in 0.8in 0in]{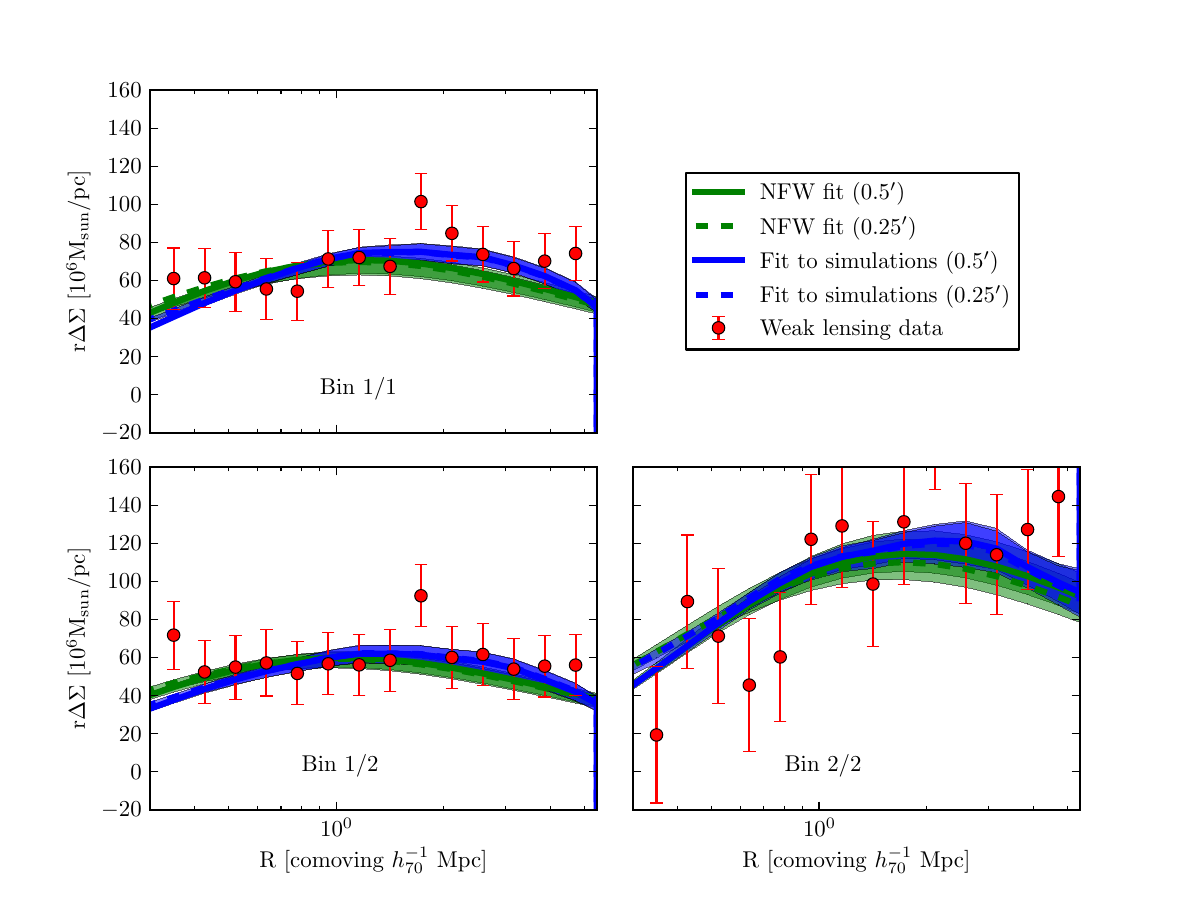}
\caption{Weak lensing measurements (in red), best-fit NFW models with $1\sigma$ error bars (in green) and best-fit interpolated simulation fit with $1\sigma$ error bars (in blue) for the RBC clusters in one or two X-ray luminosity bins.  Weak lensing errors are statistical errors; the NFW and simulation-fit error regions are defined by the 16th to 84th percentile range from the MCMC fitting procedure.  Parameters for the fits are shown in Table~\ref{bintable}.}\label{fig-fit}
\end{figure*}

Next, we fit the weak lensing signal with a simple spherically-symmetric NFW profile, miscentered using the distribution in Eq.~\ref{eq:mis_cent} in redshift slices as described in Section \ref{sec:mis_cent}.  We use the radius range $0.3-3$ $h_{70}^{-1}$ Mpc for the NFW haloes; this outer range corresponds to approximately the virial radius for a cluster at the mean mass and redshift, so we should have minimal contribution from large-scale structure.  We use a Markov chain Monte Carlo analysis to perform the fit, as implemented with the \textsc{emcee} package \citep{2013PASP..125..306F}, using 100 walkers and 2000 steps, excluding 50 steps as burn-in.  We use an MCMC here because the likelihood surface is non-Gaussian, particularly for the simulation fits described below.  We report the median of all the values in the chain as our best-fit value, and the errors are given by the distance from the median to the 16th and 84th percentile of the chains.
We show results from two different options for the width of the miscentering distribution, $0.25\arcmin$ and $0.5\arcmin$. Both fit the data relatively well, with some bins and fits preferring the lower offset and some the higher.  Since we are eliding other possible sources of error as well (such as possible insufficient modeling of the brightest cluster galaxy), we believe that these results represent the allowable range given our knowledge of these systems.  Offsets of $1\arcmin$ were tried and ruled out by increased $\chi^2$ values, and since $0.25\arcmin$ was already too narrow for some bins we did not try lower widths.  Given the insensitivity of our masses to this parameter and the fact that a reasonably large range of miscentering widths provides a good fit to the data, we do not attempt to constrain this parameter further.
Then we fix the mass-concentration relation to the relation from \citet{2013ApJ...766...32B} as implemented by the Colossus package \citep{2015ApJ...799..108D}, which gives us a typical concentration of $c_{200} \sim 3.5-4.5$ for this mass range. The best-fitting NFW profile, and the range of fits comprising the $\pm 1\sigma$ masses, are shown in Fig.~\ref{fig-fit}. Parameters of the fit are shown in Table~\ref{bintable}.

Finally, we fit mass profiles based on our simulations, using a radius range of $0.3-5.5$ $h_{70}^{-1}$ Mpc, which is about twice the virial radius.  Since our simulations include large-scale structure, we use the increased radial range for increased statistical power; the actual mass values we find differ by only a fraction of the $1\sigma$ error bars compared to the values we get for an upper radius limit of 3 $h_{70}^{-1}$ Mpc, so we are confident this does not bias our final results.  We produce average $\Delta\Sigma$ profiles for a set of mass bins extending $0.60$ in base-ten log mass space, with endpoints spaced every $0.24$ in base-ten log space, to construct a set of overlapping mass bins containing a distribution of cluster masses.  We then linearly interpolate between each neighboring, overlapping pair of mass bins to produce a $\Delta\Sigma$ profile at intermediate masses, and spline interpolate in the radial direction to produce a smooth function.  We fit for the interpolation parameters: the index of the lower bin $n$ and the fraction $f$ needed to produce the best-fit $\Delta\Sigma$, such that
\begin{equation}
\Delta\Sigma_{\mathrm{data}} = (1-f)\Delta\Sigma_n + f\Delta\Sigma_{n+1}
\end{equation}
We then interpret the mean mass of the fit to be the corresponding linear combination of the mean masses of the clusters in mass bins $n$ and $n+1$.

In principle, as the $\Delta\Sigma$ profiles are a power of the mass, our linear combination should properly be a logarithmic mean.  However, in practice, our bins are finely spaced enough, and the curves noisy enough, that we can interpolate linearly without affecting the precision of our results.  We tested this by generating mass bins from the simulations with the same size as the bins used for fitting the mass, but with offsets from the endpoints of those mass bins.  With both linear and logarithmic interpolation, we find that at low to intermediate masses our fitting routine returns the expected answer.  The relation breaks down above $\sim 7 \times 10^{14} h_{70}^{-1} M_\odot$, where our fitting code begins to return systematically low masses. The discrepancy rapidly increases with true mass, reaching a bias of 20 per cent or greater at masses greater than $\sim 8 \times 10^{14} h_{70}^{-1} M_\odot$.  This breakdown above $\sim 7 \times 10^{14} h_{70}^{-1} M_\odot$ corresponds to approximately the most massive 10 per cent of our sample (10-15 objects, depending on where we draw the line between stochastic noise and systematic bias).  This is close to the values found for the higher-mass bin of the sample when it is split into two bins, and is possibly a source of error for those points, but we do not expect this to be a dominant source of error even for this bin (an expectation held up by the general consistency of the results with the other bins and fitting methods).

\begin{figure}
\includegraphics[width=0.45\textwidth]{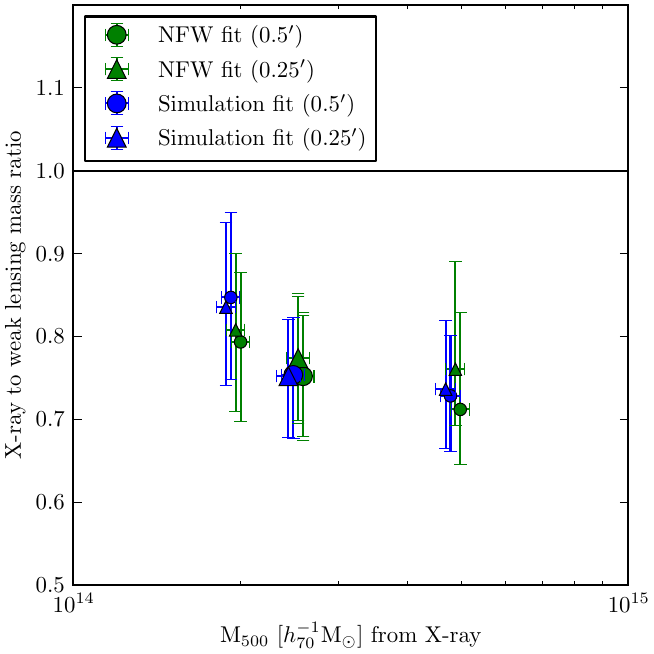}
\caption{Mass ratios from the cluster bins of Table \ref{bintable} for NFW profile fits (green) and interpolated simulation predictions (blue), plotted against a weighted average of the X-ray luminosity mass for each bin, slightly offset in the x-direction for clarity.  Small points are the results from the 2-bin case, large points from the 1-bin case.  Error bars are $1\sigma$ statistical errors from the fitting procedure on the y-axis and scatter within the X-ray bin on the x-axis.  The black line is unity.}\label{fig-masscomparison}
\end{figure}

The interpolated curves and their error region are shown in green in Fig. \ref{fig-masscomparison}.  The interpolated simulation fits have a lower (or approximately equal) $\chi^2$/dof for almost all fits, except the less massive of the two bins in the two-bin case for one value of the miscentering offset.  By eye, the NFW profile is clearly ``peakier'' than the interpolated simulation fits, which we expect since baryonic effects in the hydrodynamic simulation make the cluster mass profiles less cuspy \citep[e.g.][]{Duffy2010}.  
The data appears more consistent with the simulation trend at small radii, though not to a statistically significant level.  This effect is not strongly degenerate with the miscentering effect, which tends to move the peak in radius, but not broaden it sufficiently to overcome the discrepancy in the small-scale behavior of the density distribution.  We did not investigate altering the mass-concentration relation of our NFWs to better fit the simulation results, as the simulations look non-NFW-like in other ways as well.

A number of corrections are necessary to go from the raw X-ray averages and the raw lensing data to the mass comparison that is our final result.  First, the lensing signal must be corrected for the known errors in our photo-$z$ catalogue, as defined in section \ref{WLphotoz}. After we fit our parametric models (either simulation or analytic NFW) to the data, we obtain a ``mass'' which is somewhere between the median and the mean of the actual distribution of masses that went into the signal, since the lensing signals are not linearly proportional to the mass; we must define an additional correction to take our fitted masses and turn them into real mean masses.  For the NFW case, we simulated a signal made of perfect NFW haloes with the mass distribution of the RBC clusters, with several different multiplicative biases to mimic the bias effect, and compared our fit values to the known mean mass to calibrate this effect separately from errors caused by profile differences.  The results were very stable across bias values covering the range found in this work.  The largest correction is for the single mass bin case, as expected, with an 11 per cent calibration factor, while the two-bin case has 6 per cent and 2 per cent for the lower-mass and higher-mass bins respectively.  For the simulation case, we fit the predictions shown in Fig.~\ref{fig-simpredict} with our interpolation scheme and use the resulting bias as our correction, which is 6 per cent for bin 1/1, 0.1 per cent for bin 1/2, and 8.5 per cent for bin 2/2.  Finally, we must weight the X-ray masses from the RBC catalogue by the expected geometric lensing factors (nearby clusters take up more area on the sky, meaning more lens-source pairs in the sums, and also higher lensing efficiency due to the distance ratios involved) and correct the resulting average for Malmquist bias.  We use the relationship between true mass and ``observed'' (scattered) mass in our simulations to measure the Malmquist bias correction factor, while the geometric lensing weights are purely analytic once we fix the cosmology to our fiducial model.  The corrections range from 0 to 5 per cent for photo-$z$; from 0 to 11 per cent for lensing mass vs mean mass differences; and from 9 to 15 per cent for Malmquist bias.

We note that, while both the X-ray and weak lensing masses are reported at $\Delta_c=500$, the weak lensing masses are the product of a parametric fit that goes to larger radius, so the masses are not measured on the same aperture.  

Finally, in Figure \ref{fig-masscomparison}, we show the two mass determinations (NFW and simulation) for the two miscentering parameters in comparison to the expected average mass derived from the X-ray luminosity masses.  The simulation fits and NFW fits show a clear trend for a larger weak lensing mass than X-ray mass, with a size of approximately 25 per cent, consistent with our expectations from simulations. 

\subsection{Sources of systematic error}

Our errors in the above figures are statistical only.  How much might systematic errors affect our results?  We consider three sources of error: errors from problems with the source catalogue, errors from problems with the lens catalogue, and modeling errors.

The source catalogue errors have been well-characterized in a series of papers
  \citep{2012MNRAS.420.1518M,2012MNRAS.425.2610R,2012MNRAS.420.3240N,2013MNRAS.432.1544M} and come
  from a combination of shear estimation errors and photometric redshift errors.  A detailed
  breakdown of the systematic error budget is given in \citet{2013MNRAS.432.1544M}; for this
  catalogue, we expect the systematic uncertainty on our $\Delta\Sigma$ measurement to look like a
  top hat with width 4 per cent, or--in a Gaussian frame--a Gaussian error with width 2 per cent.
  As $\Delta\Sigma$ is approximately proportional to the mass to the $2/3$ power in the inner
  regions of NFW haloes \citep{2010MNRAS.405.2078M}, this results in a 3.5 per cent mass uncertainty due to our shape and photometric redshift catalogue.

Problems with the lens catalogue consist mainly of objects identified as clusters that are not, in fact, real clusters.  Given the flux limits we use to restrict the RBC clusters, we expect that the sample is approximately 100 per cent pure \citep{Mantz2010}.  Some cluster X-ray measurements are also likely contaminated by AGN emission, but this should be addressed by our bias and scatter calculations, so we do not add a further item here.

Finally, we turn to modeling errors.  We might see errors from models that have the wrong
  radial $\Delta\Sigma$ profile; this difference is captured by the differences between our theoretical NFWs and our simulation fits, and is of order 2 per cent for Bins 1/1 and 1/2 and 3 per cent for Bin 2/2.  Incorrect miscentering would also impact our results; as we see differences of 2 per cent in our analysis with different miscentering widths (6 per cent for Bin 2/2), we would, again, expect that our errors from this parameter are of the same order.  Our models also implicitly include the scatter within the bin: we correct for this in the theoretical models and include it directly in the simulation models.  If we are incorrect about the size of the scatter, we will be wrong about the size of this correction.  This does not impact Bin 1/1 very much (changing the scatter by 25 per cent up or down changes the correction by a per cent or so); for the bins split by mass, meaning a larger fraction of clusters impacted by a sharp cutoff at the bin edge, the correction is larger, 3 per cent for Bin 1/2 and 10-20 per cent for Bin 2/2.  These systematic error bars are quite large for Bin 2/2, which we believe is a result of the rarity of those objects: few such clusters can be found in the simulation volume, and there may also be scatter in our accounting of the correction for the scatter that we apply to the theoretical NFWs.  The systematic errors for the upper bin, then, reflect additional stochasticity from the small number of rare objects.  There are additional potential sources of error, but we believe these dominate our systematic error budget for this work.

In total, we have: shape and photo-$z$ catalogue errors of 3.5 per cent; errors from incorrect model shape of 2 per cent (3 per
  cent for Bin 2/2); errors from miscentering of 2 per cent (6 per cent for Bin 2/2); and errors
  from potential misunderstood scatter (1 per cent for Bin 1/1, 3 per cent for Bin 1/2, and 15 per
  cent for Bin 2/2). Added in quadrature, since these are largely independent uncertainties, this results in a 5 per cent systematic error bar for Bins 1/1 and 1/2 and a 17 per cent systematic error bar for Bin 2/2.

\section{Conclusion}\label{Conclusion}

We have used weak lensing measurements to calibrate the average masses for 166 clusters with X-ray measurements in the RBC catalogue. We find evidence that the X-ray masses from scaled luminosities in the RBC catalogue are approximately 15 to 30 per cent lower than the weak lensing masses over the range of masses probed in this cluster sample (0.6 to $10.5\times 10^{14} h_{70}^{-1} M_\odot$).  We lack sufficient statistical power to address whether the bias is mass-dependent.  We cannot say with certainty what fraction of this mass bias is due to the assumption of hydrostatic equilibrium: given the discrepancy in X-ray masses between different analysis methods \citep{2014MNRAS.438...49R}, it is likely that some of the bias is due to effects other than the assumption of hydrostatic equilibrium, such as differing X-ray calibrations, errors in the original $L_{\X}-M$ fit in addition to problems of hydrostatic equilibrium, and discrepancies between X-ray and weak lensing estimates of $R_{500}$.  However, our analysis is agnostic to the sources of difference between the X-ray and weak lensing estimates, and merely produces a corrected version of the masses given by scaled luminosity in he RBC catalogue. This measurement is in agreement with the expectations from simulations \citep[e.g.][]{2014arXiv1406.3451O,2014arXiv1405.7876D,2014MNRAS.440.2077M}. 

To estimate the basic implications for a cosmological measurement using cluster abundances, we use the mass function from \cite{2008ApJ...688..709T} at $z=0$ and the cosmological parameters from \cite{2014A&A...571A..16P}.  Assuming use of all clusters down to a (real) mass limit of $10^{14}h^{-1}M_\odot$, then if a mass bias corresponding to $q=0.78$ is ignored, a single-parameter cosmological constraint on $\sigma_8$ would be biased low by 12 per cent.  For higher masses, this rapidly becomes worse due to the steepening of the mass function.  Obviously a real analysis would not vary only a single parameter, but this gives a rough indication of the magnitude of the problem.  Since many cosmological analyses have smaller statistical errors than this, it will be important to apply corrections for this mass bias rather than using X-ray-inferred masses directly.  We defer a full analysis of the effect on multi-parameter cosmological constraints to future work.

There are cosmological implications for this measurement with respect the previous Planck cluster cosmology analysis. In \citet{plnk2013} the assumed value for the mass bias value is consistent with our result. The masses in \citet{plnk2013} are calibrated by the REXCESS X-ray masses \citep{plnk2011} just like the MCXC and RBC catalogues. Our results are marginally consistent with the 40-50 per cent mass bias required to bring the cosmological parameters from the Planck cluster analysis into agreement with the primary CMB, and agree well with previous analyses such as \citet{2014MNRAS.443.1973V} (30 per cent bias) and \citet{2015MNRAS.449..685H} (25 per cent bias).  Our analysis does not rule out contributions for other systematics in the Y-M relationship used in \citet{plnk2013}.
	
Our reported mass bias of $25 \pm 7$ stat. $\pm 5$ sys. per cent is
within $2\sigma$ of the recent results from the LoCuSS subsample
presented in \citet{2015arXiv151101919S} from mass measurements in
\citet{2015arXiv150704493O}. They find a $5 \pm 5$ per cent
hydrostatic mass bias.
When comparing these results one should keep two things in mind: the
X-ray masses use different scaling relations and our results and the
results of \citealt{2015arXiv150704493O} apply to different cluster
samples. In relation to the Planck cluster cosmology results, the bias
we calculate needs to be carefully propagated through the previous
Planck calibrations of cluster masses and the
\citet{2015arXiv151101919S} bias should be carefully corrected for the
Eddington bias inherent in the Planck cluster masses
\citep{2015arXiv150908930B}.

Beyond the Planck cluster analysis our results are an important calibration for any results that use the RBC cluster mass estimates in their cosmological parameter estimation. We have quantified a systematic bias that other previous analyses \citep[e.g.][]{HB2013} had to marginalize over with broad priors. New or reanalyses of measurements that use the RBC cluster catalogue will significantly improve with our mass calibration.

\section*{Acknowledgements}

MS and RM acknowledge the support of the Department of Energy Early Career Award program.   NB acknowledges the support from the Lyman Spitzer Fellowship.  US is supported in part by the NASA ATP grant
NNX12AG71G. 

The authors thank Eduardo Rozo for helpful comments on the draft of this paper and the anonymous referee for their detailed and thoughtful suggestions.

\bibliographystyle{mnras}
\bibliography{RBC_WL,wl_papers}

\end{document}